\documentclass[conference]{IEEEtran}
\IEEEoverridecommandlockouts
\ifCLASSINFOpdf
\else
\fi
\usepackage{textcomp}
\usepackage{graphicx}
\usepackage{float}
\usepackage{cite}
\usepackage{booktabs}
\usepackage{epstopdf}
\usepackage{multirow}
\usepackage{boldline}
\usepackage{amsmath}
\usepackage{relsize}
\usepackage{array}
\usepackage{makecell}
\usepackage{colortbl}
\usepackage{algorithm}
\usepackage{algpseudocode}
\usepackage[normalem]{ulem}

\def\infinity{\rotatebox{90}{8}}

\makeatletter
\newlength \figwidth
\if@twocolumn
\else
\fi
\makeatother
\begin{document}
\IEEEoverridecommandlockouts
%
\title{Cross-layer optimized routing with low duty cycle TDMA across multiple wireless body area networks}
\author{\IEEEauthorblockN{Samiya~Shimly}
\IEEEauthorblockA{The Australian National University$^\mathsection$\\
Data61$^\dagger$, CSIRO\\
Email: Samiya.Shimly@data61.csiro.au}
\and
\IEEEauthorblockN{David~B.~Smith}
\IEEEauthorblockA{Data61$^\dagger$, CSIRO\\
The Australian National University\\
David.Smith@data61.csiro.au}
\and
\IEEEauthorblockN{Samaneh~Movassaghi}
\IEEEauthorblockA{The Australian National University$^\mathsection$\\
Data61$^\dagger$, CSIRO\\
Samaneh.Movassaghi@data61.csiro.au}\vspace{-50pt}
\thanks{$^\dagger$National Information and Communications Technology Australia (NICTA) has been incorporated into Data61 of CSIRO. $^\mathsection$This research is supported by an Australian Government Research Training Program (RTP) scholarship.}}

%

\maketitle
\begin{abstract}
In this paper, we study the performance of two cross-layer optimized dynamic routing techniques for radio interference mitigation across multiple coexisting wireless body area networks (BANs), based on real-life measurements. At the network layer, the best route is selected according to channel state information from the physical layer, associated with low duty cycle TDMA at the MAC layer. The routing techniques (i.e., shortest path routing (SPR), and novel cooperative multi-path routing (CMR) incorporating 3-branch selection combining) perform real-time and reliable data transfer across BANs operating near the 2.4 GHz ISM band. An open-access experimental dataset of `everyday' mixed-activities is used for analyzing the proposed cross-layer optimization. We show that CMR gains up to 14 dB improvement with 8.3\% TDMA duty cycle, and even 10 dB improvement with 0.2\% TDMA duty cycle over SPR, at 10\% outage probability at a realistic signal-to-interference-plus-noise ratio (SINR). Acceptable packet delivery ratios (PDR) and spectral efficiencies are obtained from SPR and CMR with reasonably sensitive receivers across a range of TDMA low duty cycles, with up to 9 dB improvement of CMR over SPR at 90\% PDR. The distribution fits for received SINR through routing are also derived and validated with theoretical analysis.
\end{abstract}
%
\IEEEpeerreviewmaketitle
\section{Introduction}
Wireless body area networks (BANs) are often specifically designed for healthcare scenarios to autonomously connect various medical sensors and actuators located on, in, around or/and near the human body to monitor physiological signals. The IEEE $802.15.6$ BAN Standard aims to enable low-power communication to be reliable and practical for in-body/on-body nodes to serve a variety of medical and non-medical applications \cite{tg6_std}. With the anticipated growth in the number of people using BANs, their coexistence is a concern for the near future, where reliable communications is vital in healthcare scenarios particularly. When multiple closely-located BANs coexist, the potential inter-network communication and cooperation across BANs leads to the implementation of wireless body-to-body networks (BBNs) \cite{meharouech2015future}. The main motivation behind BBN is to make use of body-to-body (B2B) communication to overcome the problems of coexistence and general performance degradation for closely located BANs. This type of network could provide cost-effective solutions for remote monitoring of a group of patients, for instance, by relaying physiological data in case of out-of-range network infrastructure.

A number of interference-aware coexistence schemes for multiple BANs have been proposed in\cite{dong2013opportunistic, movassaghi2016enabling}. A Cross-layer Opportunistic MAC/Routing protocol (COMR) has been proposed in \cite{abbasicross} for improving reliability in BAN, where the authors have used a timer-based approach with combined metrics of residual energy and RSSI as their relay selection mechanism in a single BAN. In \cite{tseng2016efficient}, the authors have proposed an efficient cross-layer reliable retransmission scheme (CL-RRS) without additional control overheads between the physical (PHY) and MAC layer, which significantly improves frame loss rate and average transmission time as well as reduces the power consumption. However, most previous works have not considered practical BAN coexistence, using actual measured data, for intra-BAN and inter-BAN communications in tiered architectures.

In this paper, we perform cross-layer optimization across the physical, MAC and network layers for two-tiered communications, with on-body BAN at the lower tier and BBN at the upper tier. Time division multiple access (TDMA) is used as the MAC layer protocol with low duty cycling for improving co-channel interference. Among the two top foremost popular medium access techniques, i.e., TDMA and CSMA/CA (carrier sense multiple access with collision avoidance) used in BANs, TDMA has maximum bandwidth utilization and lower power consumption compared to CSMA/CA \cite{filipe2015wireless}. The analysis is applied to an open-access radio measurement dataset provided in\cite{smith2012body} (captured using NICTA$^\dagger$ developed wearable channel sounders/radios), recorded from `everyday' mixed-activities and a range of measurement scenarios with people wearing radios. Our key findings in this paper, based on empirical data obtained from real-life measurements are as follows:
\begin{itemize}
    \item The proposed CMR obtains up to $14$ dB and $10$ dB improvement over SPR  with higher ($8.3\%$) and lower ($0.2\%$) duty cycles, respectively, at $10\%$ outage probability with respect to an acceptable SINR in a dynamic environment associated with mobile subjects.
     \item With $90\%$ packet delivery ratio (PDR), CMR provides up to $9$ dB (with $8.3\%$ duty cycle) and $8$ dB (with $0.2\%$ duty cycle) performance improvement over SPR, at $-89$ dBm receive sensitivity.
     \item In the best-case scenario (at $-100$ dBm receive sensitivity), both SPR and CMR achieve almost $100\%$ PDR (equivalent to negligible packet error rate).
    \item CMR is more spectrally efficient than SPR with a spectral efficiency of up to $0.15$ bits/s/Hz at $-95$ dBm receive sensitivity.
    \item The empirical received SINR through SPR has an inverse gaussian or, lognormal distribution while the empirical received SINR through CMR has a Burr (type XII) distribution.
\end{itemize}
The cross-layer methods described in this paper can incorporate both postural body movements (intra-BAN communication) and mobility (inter-BAN communication) together with interference-aware routing and excellent communication reliability across coexisting BANs.

\section{System Model}
\begin{figure}[!t]
\centering{\includegraphics [width=2.5in]{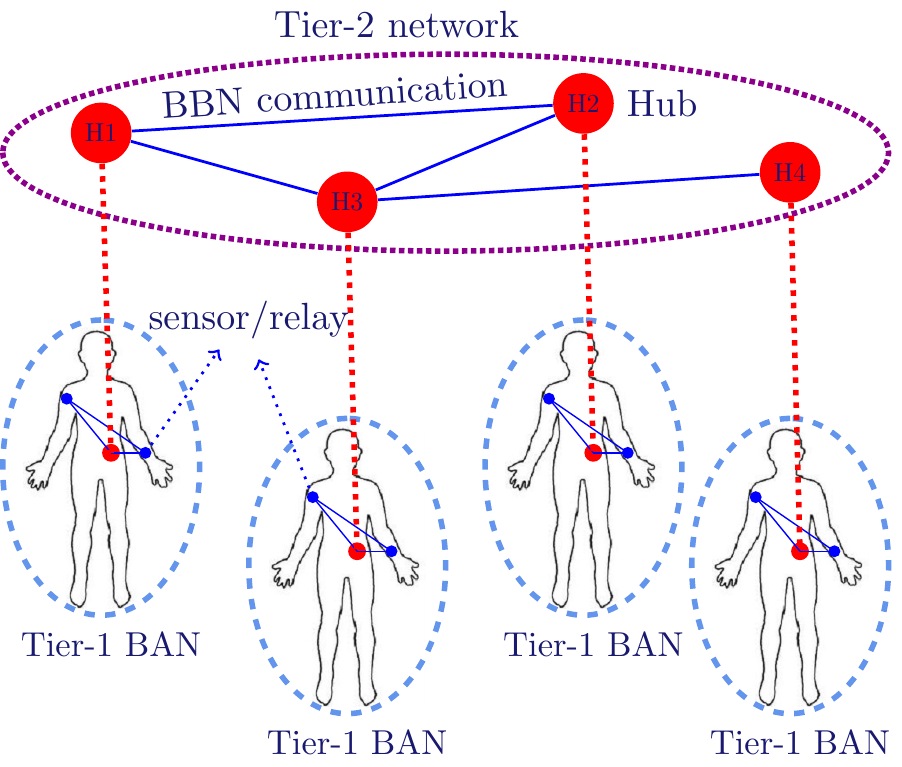}}
\caption{Two-tiered architecture of $4$ coordinated BANs}
\label{Tier}
\end{figure}

In this work, $10$ co-located mobile BANs (people with fitted wearable radios) are deployed for experimental measurements where some of the BANs are considered as coordinated and the others are causing interference by coming in the range of the coordinated BANs. We discuss the performance of $4$ coordinated BANs where $6$ interfering BANs coming in the range of the transmissions. We assume a two-tiered network architecture formed from the coordinated BANs, where the hubs of the BANs are in tier-$2$ in a mesh (inter-BAN/ BBN communication) and the on-body sensors of the corresponding BANs are in tier-$1$ (intra-BAN communication). An abstraction of the architecture is given in Fig. \ref{Tier}, with four coordinated BANs. It can be portrayed as a hybrid mesh architecture where BANs (hubs/gateways) are performing as both clients and routers/relays, which will enable flexible and fast deployment of BANs to provide greater radio coverage, scalability and mobility.

Dynamic routing is performed at network layer in a cross-layered approach, with two different routing techniques, i.e., shortest path routing (SPR) and cooperative multi-path routing (CMR), that utilize and interact with the physical and MAC layers. Therefore, changes in channel states are directly indicated from the physical layer to the network layer, so that the routes with most favorable channel conditions are chosen. TDMA is employed as the co-channel access scheme across all BANs to enable co-channel interference mitigation. As global coordination is not feasible across coexisting BANs \cite{dong2013opportunistic}, the starting time of each coordinated node is randomly selected from an uniform distribution between $0$ ms and the idle period $([0, T_{idle}])$. The idle period of the coordinated nodes, $T_{idle}$ is calculated as $T_{idle} = (n_d-1)(P_{trans}\times t)$, where $n_d$ is the total number of nodes (hubs + relays/sensors) of the coordinated BANs, $P_{trans}$ is the number of packets transmitted per node and $t$ is the packet transmission time. The duty cycle, $D_c$ of a given node is measured as follows:
\begin{equation}\label{dc}
D_c = \Big(\frac{P_{trans} \times t}{\Delta}\Big) \times 100\%
\end{equation}
where $(P_{trans} \times t)$ is the active period of the node in a given time period $\Delta$.

To substantially decrease the interference level, the duty cycles are lowered by increasing the idle period of the nodes, hence decreasing the active period. The timestamp\footnote{The routing table updates after each timestamp and the samples are taken periodically over the timestamp period.} is set to be $10$ times the sampling period of the coordinated BANs for considering different ranges of duty cycles.
The signal-to-interference-plus-noise ratio (SINR) for any given link/branch is measured as follows:
\begin{equation}\label{sinr}
\gamma_s(\tau) = \frac{p_{tx} \left | h_{s,d}(\tau)\right | ^2}{\mathlarger{\sum}_{i=1}^{n} \big(p_{tx} \left | h_{int_i,d}(\tau)\right | ^2\big) + \left | \nu(\tau)\right | ^2}
\end{equation}
where, $\gamma_s(\tau)$ is the measured SINR value of a signal $s$ at time instant $\tau$, $p_{tx}$ is the transmit power, $n$ is the number of interfering nodes, $\left | h_{s,d}\right |$ and $\left | h_{int_i,d}\right |$ are the average channel gains across the time instant of the signal-of-interest and the $i^{th}$ interfering signal, respectively. $\left | \nu \right |$ represents the instantaneous noise level at the destination node. The received noise power is set at $-100$ dBm.

\subsection{Experimental scenario}
The open-access dataset on which we base our analysis, consists of continuous extensive intra-BAN (on-body) and inter-BAN (body-to-body) channel gain data of around $45$ minutes, captured from $10$ closely located mobile subjects (adult male and female). 
The experimented subjects were walking together to a hotel bar, sitting there for a while and then walking back to the office. Each subject wore $1$ transmitter (hub) on the left-hip and $2$ receivers (sensors/ relays) on the left-wrist and  right-upper-arm, respectively (Fig. \ref{Tier}). The radios were transmitting at $0$ dBm transmit power with $-100$ dBm receive sensitivity. A description of these wearable radios can be found in \cite{hanlen2010open} and the experimental dataset can be downloaded from \cite{smith2012body}. Each transmitter transmits in a round-robin fashion, at $2.36$ GHz, with $5$ ms separation between each other. For example, with $4$ coordinated BANs (a total of $12$ nodes), each transmitter is transmitting every $60$ ms to every $3$ other subject's receivers as well as their own receivers (all small body-worn radios/hubs/sensors), along with capturing the RSSI (Receive Signal Strength Indicator) values in dBm. For real-time dynamic estimation, we timestamp the samples of a given link periodically with a continuous timestamp period of $600$ ms, given the longer coherence times of up to $1$ s for the `everyday' mixed activity for on-body narrowband BAN channels \cite{smith2013propagation}. Due to the reciprocity property, the channel from any $T_x$ (transmitter) at position $a$ to $R_x$ (receiver) at position $b$ is similar for $T_x$ at $b$ to $R_x$ at $a$ \cite{hanlen2010open}, thus transmitters and receivers can be considered interchangeably.
\subsection{Proposed routing approach}
The notion of BBN is more dynamic and potentially large-scale, where each BAN can join and/or leave the network seamlessly, without the need for any centralized infrastructure. Hence, dynamic routing is necessary to enable routers to select paths according to real-time logical network layout changes  by periodic or on-demand exchange of routing information. Here, we have implemented dynamic routing based on Open Shortest Path First (OSPF) protocol, which uses Dijkstra's algorithm.
\subsubsection{Dynamic Shortest path routing}
We perform dynamic shortest path routing (SPR)\cite{wang1992analysis} with experimental measurements based on link-state algorithm, where the source nodes intend to find routes with a minimum cost (based on routing metrics) to their destinations and update the routing table dynamically to adapt variable channel conditions and topological changes. We use a combination of two routing metrics: ETX (Expected Transmission Count) and hop count. Hop count identifies the route which has minimal number of hops. The ETX path metric is a simple, proven routing path metric that favors high capacity and reliable links. This metric estimates the number of retransmissions required to send unicast packets by measuring the loss rate of broadcasted packets between pairs of neighboring nodes \cite{malnar2009comparison}, which can be calculated as follows:
\begin{equation}\label{etx}
ETX = (1-O_p)^{-1}
\end{equation}
where $O_p$ is the outage probability.
ETX adds more reasonable behavior under real life conditions, since this metric is based on packet loss and thus the number of packets sent. In this paper, an optimal path with lowest cost possible is selected by combining these two metrics (ETX + Hop count), restricting the hop count to two hops. In SPR, the combined SINR at the destination node is measured as follows:
\begin{equation}\label{sinr_spr}
\gamma_{comb (spr)} = min\big(\gamma_{H_1},\gamma_{H_2}\big)
\end{equation}
where $\gamma_{H_1}$ and $\gamma_{H_2}$ are the SINR of the first and second hops of the shortest path from source to destination, respectively.
\subsubsection{Dynamic Cooperative multi-path routing}
Multi-path routing yields better performance than single-path routing by providing simultaneous parallel transmissions with load balancing over available resources. Hence, cooperative multi-path routing (CMR) has been considered in \cite{liu2008cooperative}. We propose a new CMR scheme that employs $3$-branch selection combining (SC) within individual route paths, incorporating shortest path routing (SPR) and improving the performance of SPR.

Here, in dynamic CMR, we use multiple alternative paths from source to destination with cooperatively combined channels in each route-hop. In this paper, route-hop refers to each hop of a route in CMR from source hub to destination hub through an intermediate BAN hub (acting as a mesh router/ relay). In each route-hop, $3$-branch cooperative SC is used (because of the advantages of $3$-branch cooperative SC in BAN communications \cite{shimly2016cooperative}), where one of the branches is the direct link and the other two branches have two link-hops. Link-hop refers to each hop from a BAN hub through on-body relays of the corresponding BAN. A given node, when acting as a relay, follows the decode-and-forward relaying scheme for which it decodes the signal and then retransmits it. The equivalent channel gain at the output of SC can be estimated as follows:
\begin{equation}\label{sc}
h_{sc}(\tau) = \max \Big\{{h_{sd}(\tau), h_{sr_1d}(\tau), h_{sr_2d}(\tau)}\Big\}
\end{equation}
where $h_{sc}(\tau)$ is the equivalent channel gain at the output of SC, at time instant $\tau$. $h_{sd}$ is the channel gain from source-to-destination (direct link), $h_{sr_id}= \min \{h_{sr_i},h_{r_i d}\}$ are the channel gains of the first and second cooperative relayed links (with two link-hops, $s$ to $r_i$, and $r_i$ to $d$) for $i=[1,2]$, respectively.

\begin{figure}[!t]
\centering{\includegraphics [width=2.2in]{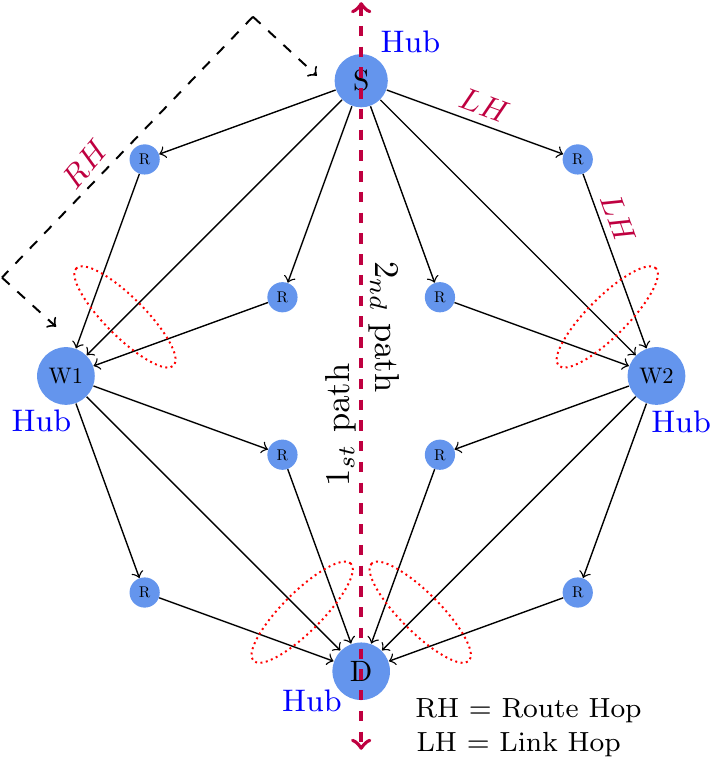}}
\caption{Cooperative multi-path routing (CMR) associated with $3$-branch selection combining}
\label{imagecmr}
\end{figure}

For multi-path routing here, two paths are used from source hub to destination hub, where both paths have two route-hops. The shortest path is chosen according to an SPR calculation. The two paths go through the two nearest BAN hubs from the source. The nearest BANs from any given source can be found from the source hub to connected hub channel gains, approximated from the RSSI at the connected BAN hubs. In CMR, the completely combined SINR at the destination is measured as follows:
\begin{equation}\label{sinr_cmr}
\gamma_{comb (cmr)} = max\big(\gamma_{P_1}, \gamma_{P_2}\big)
\end{equation}
where $\gamma_{P_i} = min\big(\gamma_{RH_1},\gamma_{RH_2}\big); [i =1,2]$ is the combined SINR of path $i$ with two route hops.
 $\gamma_{RH_j}; [j =1,2]$ is the combined SINR of the $3$-branch selection combining (by following equation (\ref{sc}) for SINR values instead of channel gains) in route hop $j$.
The process for CMR is illustrated in Fig. \ref{imagecmr} and described with a pseudocode in Algorithm \ref{alg_cmr}.

\setlength{\textfloatsep}{0pt}
\begin{algorithm}[!t]
\caption{Estimating output of CMR, incorporating SPR (with ETX + max. $2$ hops count)}\label{alg_cmr}
\begin{algorithmic}[1]
\State $\{S,D\} \gets \textit{\{Source node, Destination node\}}$
\Function{FindShortestPath}{$S$,$D$}
\State $P_{etx} \gets \textit{ETX values of every possible paths}$
\State \hspace{10mm} $\textit{from S to D}$
\State $[i , j] \gets [1 , \textit{size of } P_{etx}]$
\While{$i\ne j$}
\State $temp \gets \textit{Find min }(P_{etx})$
\If{$hop\_count(temp) = 2$}
\State $S\_to\_D \gets Path(temp)$
\Else
\State $P_{etx} \gets P_{etx} - \textit{temp}$
\State $temp \gets \textit{Find min }(P_{etx})$
\State $j \gets j - 1$
\EndIf
\State $i \gets i + 1$
\EndWhile
\If {$S\_to\_D$ is empty}
\State $S\_to\_D \gets \textit{direct path}$
\EndIf
\State \textbf{return} $S\_to\_D$
\EndFunction
\State $P_1 \gets FindShortestPath(S,D)$
\State $P_2 \gets \textit{FindShortestPath(S, D)} \notin P1$
\For {$i \gets 1,2$}
\State $P_1\_RH_i \gets selection\_combining(route\_hop_i)$
\EndFor
\State $Comb\_P_1 \gets min(P_1\_RH_1, P_1\_RH_2)$
\State $Comb\_P_2 \gets \textit{Repeat steps $24$ to $27$ for $P_2$}$
\State $\textbf{Output\_CMR} \gets max(Comb\_P_1, Comb\_P_2)$\end{algorithmic}
\end{algorithm}

\section{Performance Analysis}
\begin{table}[b!]
\centering
\caption{Applied Parameters}
\label{table_param}
\begin{tabular}{|c|c|}\Xhline{0.8pt}
Parameter & Value\\[1ex]\Xhline{0.8pt}
Bandwidth ($B$) & $1$ MHz\\\hline
Carrier Frequency & $2.36$ GHz\\\hline
Data rate & $486$ kbps\\\hline
Packet size ($\ell$) & $273$ bits\\\hline
Packet transmission time ($t$) & $0.6$ ms\\\hline
Transmit power ($p_{tx}$) & $0$ dBm\\\hline
Total Time ($T$) & $45$ mins\\\Xhline{0.8pt}
\end{tabular}
\end{table}
We consider outage probability with respect to SINR as a performance metric for the optimization techniques applied on the coordinated network in case of interference mitigation. We also estimate the packet delivery ratio and spectral efficiency of the network with respect to different receive sensitivities when applying those routing techniques on the experimental measurements. Furthermore, we investigate the theoretical results of SINR distributions produced from simulated channels (modeled with lognormal distribution with the distribution parameters found from the measured channels) and compare them with the experimental results. The applied parameters for the performance analysis are listed in Table \ref{table_param}.

\subsection{Experimental Results}
In this subsection, we discuss and compare the results found from SPR and CMR techniques with the experimental measurements. The results are averaged from $1000$ trials for obtaining comprehensive outcomes. For estimating the outages properly, the effect of non-recorded measurements (NaN) due to incorrectly decoded packets were replaced with a value of $-101$~dBm, just below the receiver sensitivity of $-100$~dBm.
\subsubsection{Outage Probability with respect to SINR}
\begin{figure}[!t]
\centering{\includegraphics[width=0.9\columnwidth]{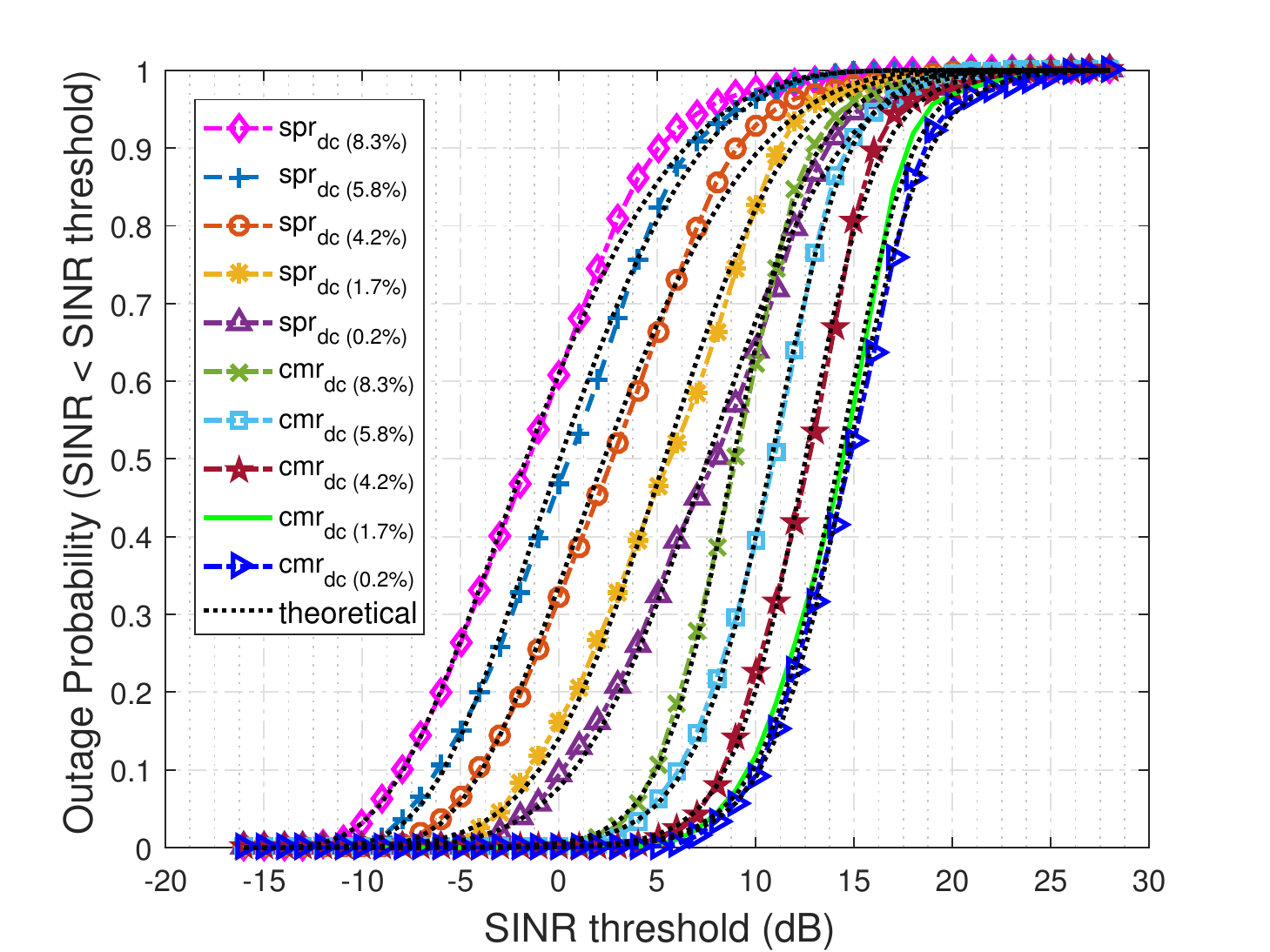}}
\caption{Average outage probability with respect to SINR threshold for SPR and CMR, with different duty cycles (dc) per node for the $4$ coordinated BANs. Receiver sensitivity $-100$ dBm, transmit power $0$ dBm; Black dotted curves represent the theoretical cdf of SINR with corresponding duty cycles}
\label{out_sinr}
\end{figure}

The outage probability with respect to SINR threshold can be expressed as,
\begin{equation}\label{outp}
P_{out} = Prob \big(\gamma_s < \gamma_{th}\big)
\end{equation}
where  $P_{out}$ is the probability of received SINR, $\gamma_s$ being less than a given threshold value $\gamma_{th}$. The averaged outage probability with respect to SINR for SPR and CMR with different duty cycles is presented in Fig. \ref{out_sinr}. As can be seen, CMR provides up to $14$ dB performance improvement over SPR at $10\%$ outage probability with respect to a SINR of $5$ dB with $8.3\%$ duty cycle. Also, with a lower duty cycle of $0.2\%$, CMR obtains up to $10$ dB performance improvement over SPR at the same outage probability with respect to a SINR of $10$ dB. Also, the best fits for the SINR distributions are validated with cumulative distribution functions (black dotted curve with each outage probability curve) where the theoretical cumulative distribution functions (cdfs) have a good match with the empirical results.

The distribution parameters found from the best fit of the SINR values (obtained from experimental measurements) from SPR and CMR with different duty cycles are given in Table \ref{table_param_dist}. According to Table \ref{table_param_dist}, the SINR values obtained from SPR provide a good fit for inverse gaussian distribution with higher duty cycles and for lognormal distribution with lower duty cycles. An inverse gaussian distribution with shape parameter $\lambda\rightarrow\infinity$ becomes more like a normal (gaussian) distribution. SINR values obtained from CMR posses a three-parameter Burr (burr type XII) or generalized log-logistic distribution. In reliability applications, the use of the log-logistic is often proposed as an alternative to the lognormal. Thus, the Burr offers an even more flexible alternative to the lognormal with all of the advantages of the log-logistic (as the log-logistic distribution is a special case of the Burr) \cite{zimmer1998burr}. The cdf of the inverse gaussian distribution is:
\begin{equation}\label{cdf_inv_gauss}
\begin{split}
F(x \mid \mu,\lambda) = \exp^{2\mu/\lambda} \Phi\Big\{-\sqrt{(\lambda/\mu)} (1+x/\mu)\Big\} + \\
\Phi\Big\{\sqrt{(\lambda x)}(x \mu - 1)\Big\}, \quad x>0
\end{split}
\end{equation}
where $\Phi$ is the cdf of the standard normal distribution, \mbox{$\mu(>0)$} and $\lambda(>0)$ are the mean and shape parameters of the inverse gaussian distribution, respectively. And, the cdf of the Burr (type XII) distribution is:
\begin{equation}\label{cdf_burr}
F(x \mid \alpha,c,k) = 1 - \Big(1+(x/a)^c\Big)^{-k}, \quad x>0
\end{equation}
where $\alpha(>0)$ is the scale parameter and $c(>0)$ and $k(>0)$ are the shape parameters of the Burr distribution. The density of the distribution is unimodal (having one clear peak) if $c>1$ and $L$-shaped if $c\leq1$.

\begin{table}[!t]
\centering
\caption{Distribution parameters of SINR values; \{$\mu$,$\lambda$\} are the mean and shape parameter of the inverse gaussian distribution; \{$\mu$,$\sigma$\} are the log-mean and log-standard deviation of the lognormal distribution; \{$\alpha$, $c$, $k$\} are the scale and two shape parameters of the burr distribution; exp. and sim. imply experimental and simulated data, respectively}
\label{table_param_dist}
\begin{tabular}{|c|c|c|c|}\Xhline{1pt}
\begin{tabular}{@{}c@{}}Routing \\ method\end{tabular} & \begin{tabular}{@{}c@{}}Duty \\ cycle\end{tabular} & \begin{tabular}{@{}c@{}}Distribution \\ fit\end{tabular} & Parameters\\[2ex]\Xhline{1pt}
\multirow{ 5}{*}{\begin{tabular}{@{}c@{}}SPR\\(exp.)\end{tabular}} & $8.3\%$ & Inv. Gaussian & $\mu = 1.799$, $\lambda = 0.511$\\\cline{2-4}
& $5.8\%$ & Inv. Gaussian & $\mu = 2.27$, $\lambda = 0.856$\\\cline{2-4}
& $4.2\%$ & Inv. Gaussian & $\mu = 4.39$, $\lambda = 1.29$\\\cline{2-4}
& $1.7\%$ & Lognormal & $\mu = 1.24$, $\sigma = 1.15$\\\cline{2-4}
& $0.2\%$ & Lognormal & $\mu = 1.74$, $\sigma = 1.22$\\\Xhline{0.8pt}
\multirow{ 5}{*}{\begin{tabular}{@{}c@{}}SPR\\(sim.)\end{tabular}} & $8.3\%$ & Inv. Gaussian & $\mu = 1.65$, $\lambda = 0.593$\\\cline{2-4}
& $5.8\%$ & Inv. Gaussian & $\mu = 2.0012$, $\lambda = 0.989$\\\cline{2-4}
& $4.2\%$ & Inv. Gaussian & $\mu = 4.1002$, $\lambda = 1.45$\\\cline{2-4}
& $1.7\%$ & Lognormal & $\mu = 1.304$, $\sigma = 1.11$\\\cline{2-4}
& $0.2\%$ & Lognormal & $\mu = 1.702$, $\sigma = 1.22$\\\Xhline{1.05pt}
\multirow{ 5}{*}{\begin{tabular}{@{}c@{}}CMR\\(exp.)\end{tabular}} & $8.3\%$ & Burr & $\alpha = 6.56$, $c = 2.58$, $k = 0.752$\\\cline{2-4}
& $5.8\%$ & Burr & $\alpha = 14.4$, $c = 2.056$, $k = 1.32$\\\cline{2-4}
& $4.2\%$ & Burr & $\alpha = 18.1$, $c = 2.35$, $k = 1.019$\\\cline{2-4}
& $1.7\%$ & Burr & $\alpha = 33.6$, $c = 2.086$, $k = 1.46$\\\cline{2-4}
& $0.2\%$ & Burr & $\alpha = 32.1$, $c = 2.105$, $k = 1.15$\\\Xhline{0.8pt}
\multirow{ 5}{*}{\begin{tabular}{@{}c@{}}CMR\\(sim.)\end{tabular}} & $8.3\%$ & Burr & $\alpha = 8.302$, $c = 2.99$, $k = 0.632$\\\cline{2-4}
& $5.8\%$ & Burr & $\alpha = 13.8$, $c = 2.49$, $k = 0.817$\\\cline{2-4}
& $4.2\%$ & Burr & $\alpha = 19.9$, $c = 2.71$, $k = 0.799$\\\cline{2-4}
& $1.7\%$ & Burr & $\alpha = 36.02$, $c = 2.26$, $k = 1.13$\\\cline{2-4}
& $0.2\%$ & Burr & $\alpha = 32.2$, $c = 2.13$, $k = 1.14$\\\Xhline{1pt}
\end{tabular}
\end{table}

\subsubsection{Packet Delivery Ratio (PDR)}
\begin{figure}[!t]
\centering{\includegraphics[width=0.85\columnwidth]{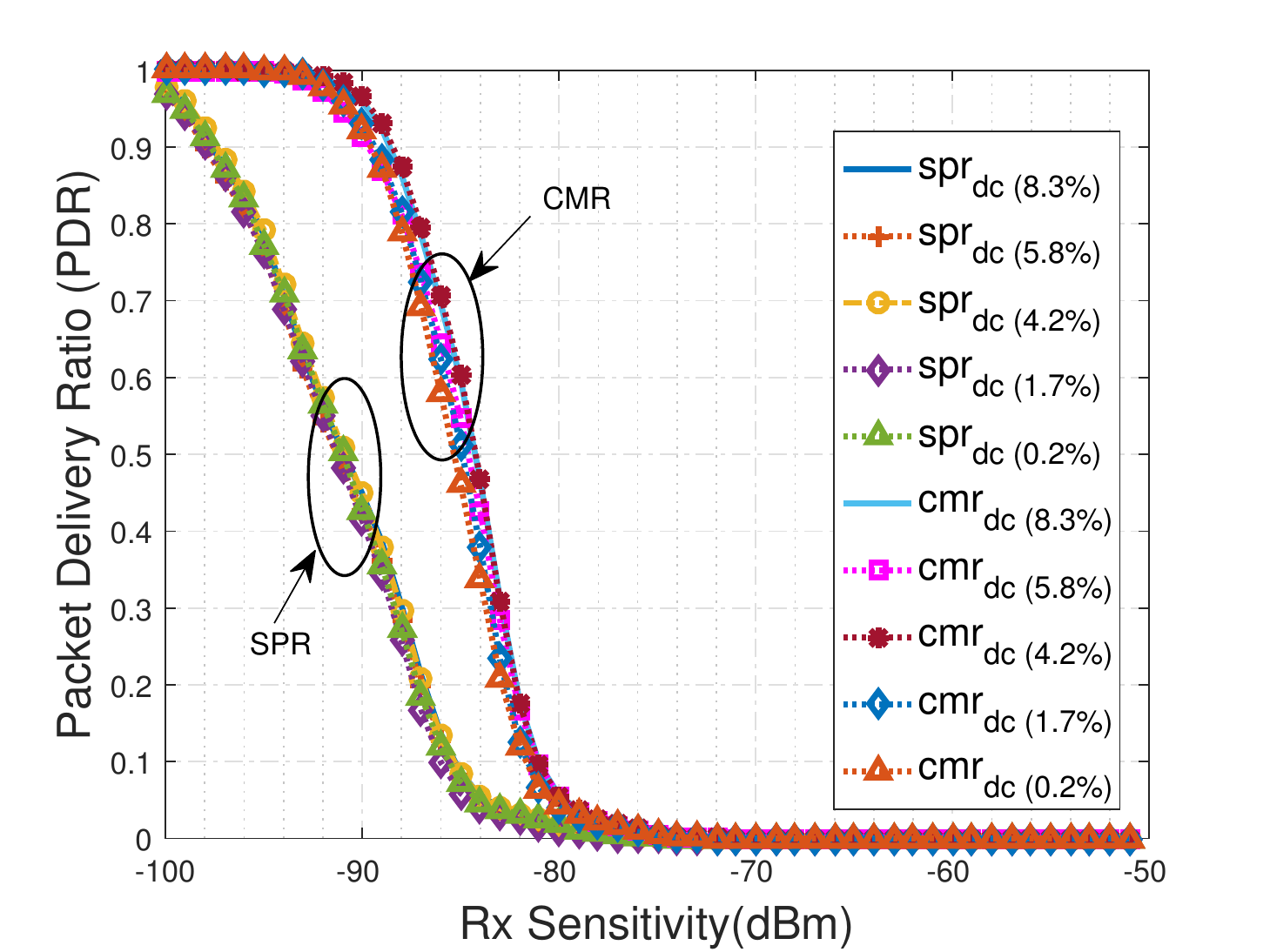}}
\caption{Average packet delivery ratio (PDR) in terms of different receive sensitivities for SPR and CMR, with different duty cycles (dc) per node for the $4$ coordinated BANs, at $0$ dBm transmit power}
\label{pdr}
\end{figure}

The packet delivery ratio (PDR) with respect to different receive sensitivities is given in Fig. \ref{pdr}. It is shown that the packet delivery ratio, which is the ratio of the successfully delivered packets ($P_{succ}$) to the transmitted packets ($P_{trans}$) at a given time, remains stable (slightly improved) with lowering the duty cycle. With a packet delivery ratio of $90\%$ (or, packet error rate (PER) of $10\%$, as $PER = 1-PDR$), the CMR provides up to $9$ dB and $8$ dB performance improvement over SPR with a higher duty cycle of $8.3\%$ and a lower duty cycle of $0.2\%$, respectively, at $-89$ dBm receive sensitivity. Also, the best-case (at $-100$ dBm receive sensitivity) PDR for SPR and CMR are almost $100\%$ which is equivalent to a negligible PER (thus fulfilling the IEEE $802.15.6$ BAN Standard requirement of PER being less than $10\%$).
\subsubsection{Spectral Efficiency}
\begin{figure}[!t]
\centering{\includegraphics[width=\figwidth]{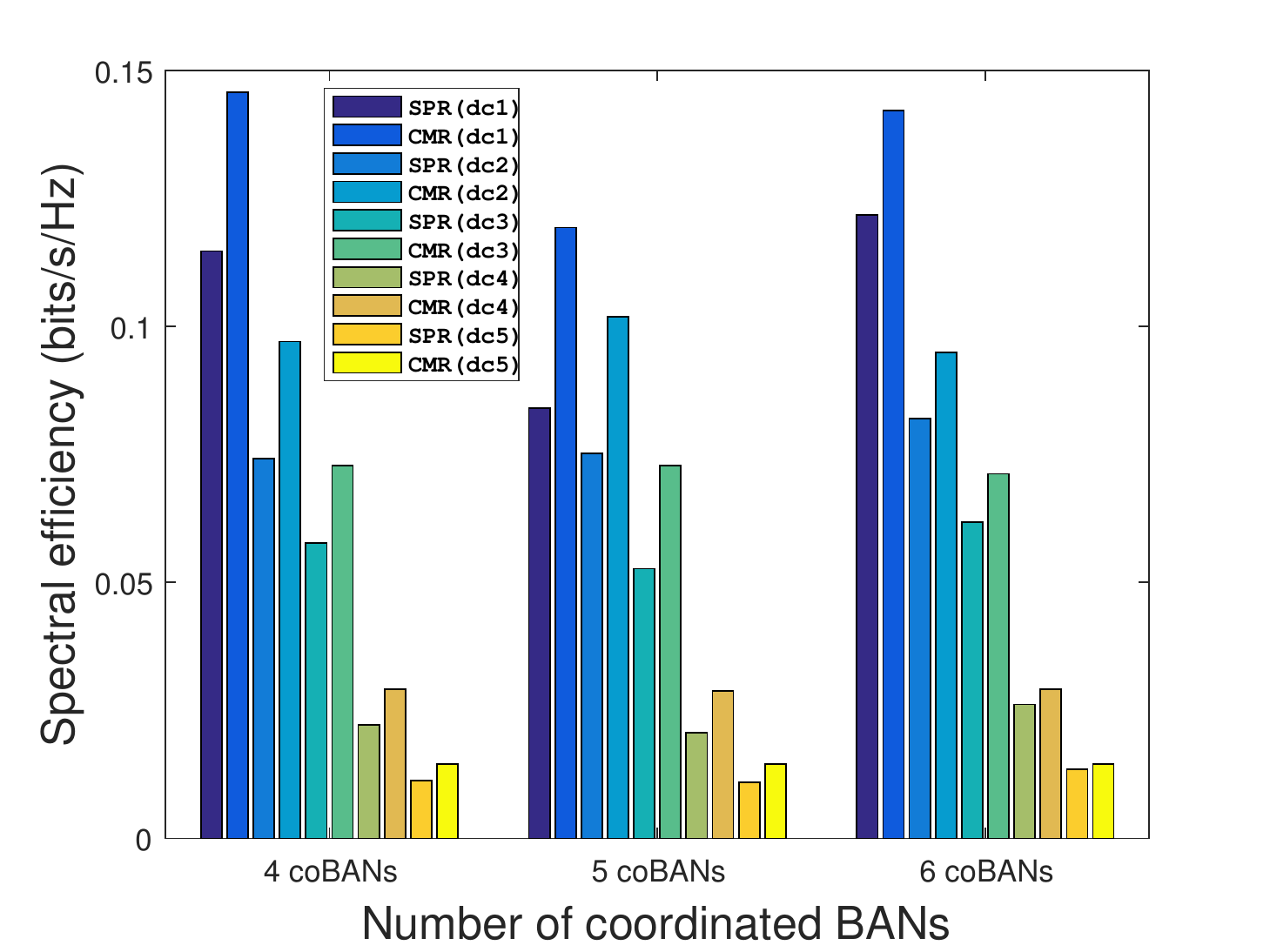}}
\caption{Average spectral efficiency with respect to $-95$ dBm receive sensitivity for SPR and CMR, with different duty cycles (dc) of $4$, $5$ and $6$ co-ordinated BANs (coBANs)}
\label{spec95}
\end{figure}

\begin{figure}[!t]
\centering{\includegraphics[width=\figwidth]{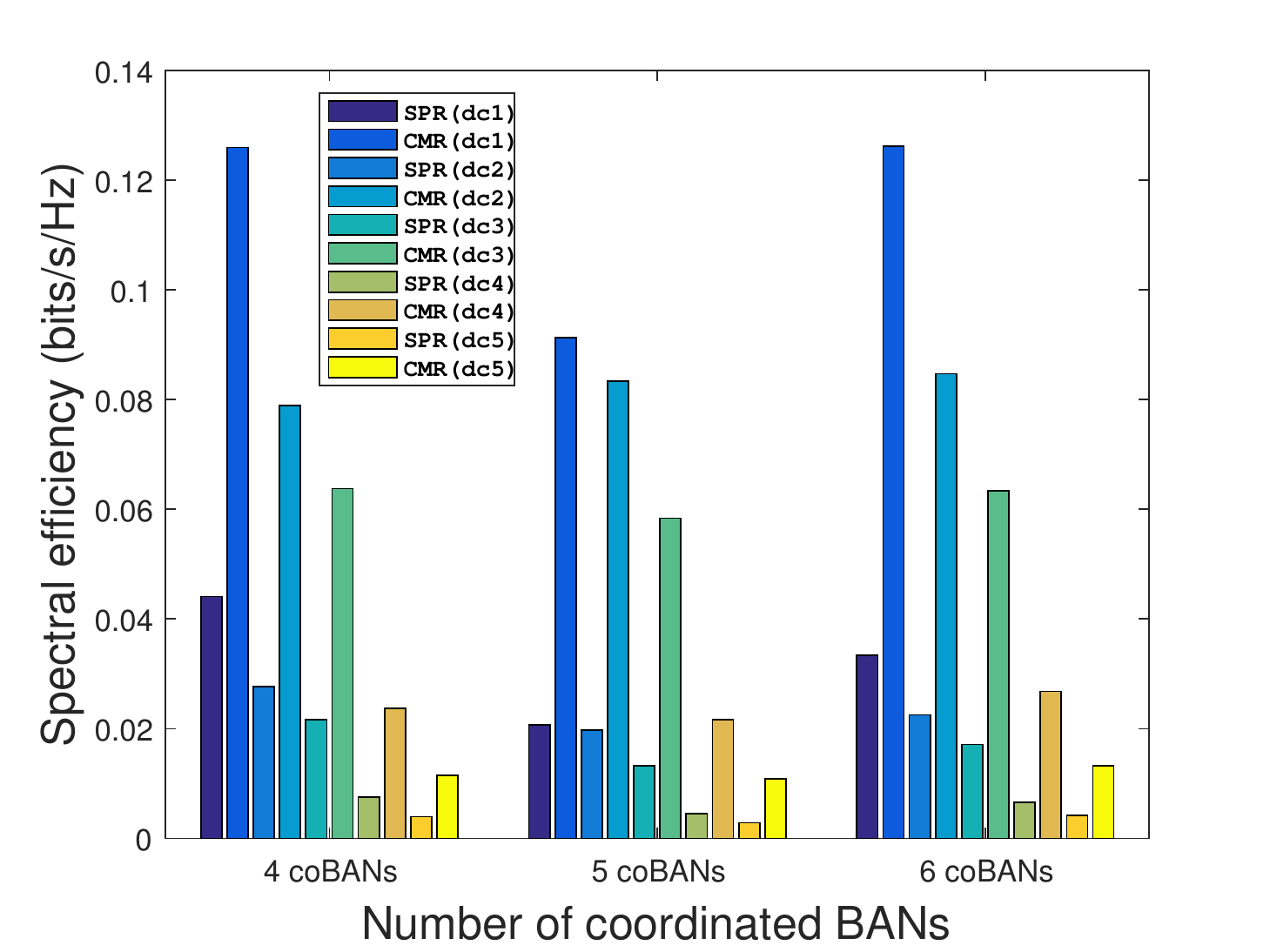}}
\caption{Average spectral efficiency with respect to $-88$ dBm receive sensitivity for SPR and CMR, with different duty cycles (dc) of $4$, $5$ and $6$ co-ordinated BANs (coBANs)}
\label{spec88}
\end{figure}

The spectral efficiency ($\zeta$) of the network with coordinated BANs is estimated as follows:
\begin{equation}\label{spef}
\zeta = \frac{\Theta \times \omega}{B}
\end{equation}
where $\omega$ is the number of coordinated BANs and $B$ is the bandwidth. The aggregated throughput of the network can be defined as $(\Theta\times\omega)$, where the throughput $\Theta$ can be measured as follows:
\begin{equation}\label{thr}
\Theta = \frac{P_{succ} \times \ell}{T}
\end{equation}
where $P_{succ}$ is the number of successfully delivered packets over the total time $T$ and $\ell$ is the length of the packet. The bandwidth and packet size can be found from Table \ref{table_param}, which are chosen in accordance with the IEEE $802.15.6$ Standard for narrowband communications \cite{tg6_std}. The average spectral efficiency with respect to different receive sensitivities (e.g. $-95$ dBm, $-88$ dBm) with different number of coordinated BANs (e.g. $4$,$5$,$6$ coBANs) and corresponding different duty cycles are presented in Figs. \ref{spec95} and \ref{spec88}. In Figs. \ref{spec95} and \ref{spec88}, duty cycles ($dc1, dc2, dc3, dc4, dc5$) refer to ($8.3\%, 5.8\%, 4.2\%, 1.7\%, 0.2\%$) for $4$ coordinated BANs, ($6.7\%, 4.7\%, 3.3\%, 1.3\%, 0.1\%$) for $5$ coordinated BANs and ($5.6\%, 3.9\%, 1.7\%, 1.1\%, 0.1\%$) for $6$ coordinated BANs. It is shown in Fig. \ref{spec95} that, CMR provides up to $0.15$ bits/s/Hz spectral efficiency with $8.3\%$ duty cycle at $-95$ dBm receive sensitivity. Also, with a lower duty cycle (e.g. $0.2\%, 0.1\%$), the spectral efficiency is greater than or equal to $0.01$ bits/s/Hz for SPR and CMR at $-95$ dBm receive sensitivity. Furthermore, Fig. \ref{spec88} shows CMR provides better spectral efficiency than SPR (with $0.2\%$ or, $0.1\%$ duty cycle, the spectral efficiency for CMR is greater than or equal to $0.01$ bits/s/Hz at $-88$ dBm receive sensitivity).

\subsection{Distributions from Simulated SINR}
For investigation purposes, we model the measured on-body and inter-body links using a lognormal distribution (as lognormal is the typical distribution for single-link narrowband small-scale fading channels \cite{smith2013propagation}). We simulate the dynamic on-body and inter-body channels according to the appropriate log-mean and log-standard deviation parameters found from the measured channels. The best fit parameters for the SINR distributions (averaged from $1000$ trials) with different duty cycles after applying SPR and CMR on the lognormally modeled channels are given in Table \ref{table_param_dist}. It can be seen that, the distribution results found from simulated channels match well with the results obtained from experimental data.

\section{Conclusion}
In this paper, we have proposed cross-layer methods, validated using experimental measurements, to optimize radio communication and mitigate interference across many co-located wireless body area networks (BANs), by utilizing distinct features at the physical, MAC and network layers. We have shown that the proposed CMR achieves up to $14$ dB performance improvement with $8.3\%$ TDMA duty cycle, and $10$ dB improvement with $0.2\%$ TDMA duty cycle over SPR, at $10\%$ outage probability with respect to an acceptable SINR. Also, CMR provides up to $9$ dB improvement over SPR with $90\%$ packet delivery ratio. Moreover, CMR contributes to suitable BAN spectral efficiency of up to $0.15$ bits/s/Hz at $-95$ dBm receive sensitivity. Thus, this work provides feasible methods (to be investigated without coordination in future) for the practical deployment of many closely-located BANs in large-scale and highly-connected healthcare systems.


%
\bibliographystyle{IEEEtran}
\bibliography{Paper_ICC_modified}

\end{document}